\documentclass{article}

\PassOptionsToPackage{numbers, compress}{natbib}
\usepackage{changepage}
\usepackage{graphicx}
\usepackage[most]{tcolorbox} 

\usepackage[final]{neurips_2024}

\usepackage[utf8]{inputenc} 
\usepackage[T1]{fontenc}    
\usepackage{hyperref}       
\usepackage{url}            
\usepackage{booktabs}       
\usepackage{amsfonts}       
\usepackage{nicefrac}       
\usepackage{microtype}      
\usepackage{xcolor}         

\title{Using AI Alignment Theory to understand the potential pitfalls of regulatory frameworks}

\author{%
  Alejandro Tlaie\\
  Ernst Str\"ungmann Institute for Neuroscience, Frankfurt am Main, Germany\\
  Technical University of Madrid, Madrid, Spain\\
  \texttt{atboria@gmail.com}
}

\begin{document}

\maketitle

\begin{abstract}
  The objective of this paper is to leverage insights from Alignment Theory (AT) research, which primarily focus on the potential pitfalls of technical alignment in Artificial Intelligence, to critically examine the European Union's Artificial Intelligence Act (EU AI Act). In the context of AT research, several key failure modes - such as proxy gaming, goal drift, reward hacking or specification gaming - have been identified. These can arise when AI systems are not properly aligned with their intended objectives. The central logic of this report is: what can we learn if we treat regulatory efforts in the same way as we treat advanced AI systems? By applying these concepts to the EU AI Act, this project uncovers potential vulnerabilities and areas for improvement in the regulation, ensuring it effectively addresses the complexities and risks associated with AI technologies.
\end{abstract}

\section{Introduction}

The advent of Artificial Intelligence (AI) technologies promises transformative changes across numerous domains, from healthcare and education to finance and law enforcement. However, the rapid deployment and integration of these technologies may also introduce profound challenges and risks, necessitating a thoughtful and robust regulatory approach. This paper seeks to critically examine the recently introduced European Union's Artificial Intelligence Act (EU AI Act) \cite{eu2024aiact} using insights derived from AI Alignment Theory (AT), a conceptual framework primarily concerned with the pitfalls of technical alignment in AI systems \cite{Ji2023AI, Gabriel2020Artificial}. AT explores various modes of misalignment in AI . These phenomena may occur when AI systems, even those engineered to high technical standards, fail to adhere to their intended ethical or operational objectives. Such misalignments can lead to unintended consequences, which, in the realm of AI, might range from inefficiencies \cite{Ji2023AI} to catastrophic risks \cite{hendrycks2023overview}, through violations of privacy \cite{Liu2021Trustworthy} or discriminatory outcomes \cite{Bohdal2023Fairness, Hoffman2020The}.

This paper proposes flipping the common logic and evaluating the EU AI Act as if it were an advanced AI system itself. By applying the principles of AT to the AI Act, the aim is to identify potential vulnerabilities and areas where the regulation might be enhanced. This methodological stance offers a unique perspective that contrasts with traditional legal analyses, focusing instead on dynamic interactions between regulatory frameworks and the technologies they intend to govern.

The EU AI Act \cite{eu2024aiact} is one of the most comprehensive regulatory efforts on achieving a safer AI, a pioneering effort aimed at creating a comprehensive legal framework for the management and oversight of AI technologies within the European Union. It categorizes AI systems according to their risk levels (see Appendix \ref{Appendix}) and sets out corresponding requirements to mitigate those risks. Despite its ambitions, there are concerns that the Act may not fully encapsulate the complexity of real-world AI applications or effectively guard against the rapid evolution of AI capabilities. 

As there already is a rich body of literature in AT, we will heavily draw on it for lenses to look through. While the focus is on four main issues identified in AT research (proxy gaming \cite{skalse2022defining}, goal drift \cite{hendrycks2024introduction}, reward hacking \cite{skalse2022defining}, or specification gaming \cite{krakovna2020specification}), the discussion remains open to additional insights that could further elucidate the strengths and weaknesses of the EU AI Act. The goal is to contribute to a deeper understanding of how such regulatory frameworks can evolve to better address the multifaceted challenges posed by AI, ensuring that they effectively safeguard public interests while promoting innovation.

In sum, this paper aims to bridge the gap between theoretical research on AI alignment and practical regulatory initiatives, particularly on AI governance. It seeks to provide a comprehensive analysis that not only highlights current limitations within the EU AI Act but also suggests pathways for future enhancements, making the regulation more resilient to the complexities and unpredictabilities of AI technology development. This exploration is intended to support policymakers, technologists, and legal scholars in their ongoing efforts to harmonize AI advancements with societal values and norms.

\section{AI systems for Education}\label{Sec2}

This section examines the current regulatory framework outlined in the AI Act, evaluating its effectiveness in addressing the complex challenges posed by AI-driven educational services. It also identifies potential areas for regulatory improvement to better protect individual rights and ensure equitable access to education. For a more detailed discussion on specific use cases and potential failure modes from AT, refer to Appendix \ref{Appendix}.

The integration of AI systems into education marks a pivotal point where technology intersects with key aspects of individual autonomy, career opportunities, and personal development. Education serves as the bedrock for accessing essential services, such as housing, loans, or health insurance, as well as for determining societal roles and responsibilities. Given the profound impact of AI in this domain, rigorous oversight is essential to prevent misuse and to uphold fairness.

\subsection{Current legal considerations}

According to the AI Act classification, deploying AI Systems for Education is considered a \textit{High risk} level. As in the case of any high risk use case, the regulation states (Art. 8-17) that AI providers must abide to the general prescriptions detailed in \ref{high_risk}. In particular, this is what the regulation says about deploying these tools in the context of Education (Art. 57):

{\centering
\begin{adjustwidth}{1cm}{1cm}
\textit{The deployment of AI systems in education is important to promote high-quality digital education and training and to allow all learners and teachers to acquire and share the necessary digital skills and competences, including media literacy, and critical thinking, to take an active part in the economy, society, and in democratic processes. However, AI systems used in education or vocational training, in particular for determining access or admission, for assigning persons to educational and vocational training institutions or programmes at all levels, for evaluating learning outcomes of persons, for assessing the appropriate level of education for an individual and materially influencing the level of education and training that individuals will receive or will be able to access or for monitoring and detecting prohibited behaviour of students during tests should be classified as high-risk AI systems, since they may determine the educational and professional course of a person’s life and therefore affect that person’s ability to secure a livelihood.}
\end{adjustwidth}}

\subsection{Looking through the AT lens}

Given that education relies on what the societal demands are and that these are, in part, shaped by the way in which citizens are educated, we think it might be informative to inspect this feedback loop through the lens of Goal Drift \cite{hendrycks2024introduction}:

\begin{tcolorbox}[colback=white, 
                 colframe=black, 
                 boxrule=0.5pt, 
                 sharp corners, 
                 title=Societal Goal Drift and misaligned AI systems] 
\textbf{Scenario:}\\
AI systems in education are initially designed to promote high-quality digital education and training, ensuring all learners acquire necessary skills such as media literacy and critical thinking. These systems aim to provide personalized educational experiences that foster all-encompassing student development. However, as the educational landscape and societal priorities evolve, there is a risk that the focus of these AI systems may drift from their original goals, potentially causing harm to students' personal development.
\end{tcolorbox}

The outlined scenario can plausibly lead to the following sequence:

\begin{enumerate}
    \item An AI system is implemented across various educational institutions to personalize learning experiences, identify individual strengths and weaknesses, and provide tailored educational pathways. Initially, this system helps educators by suggesting diverse and inclusive curricula that promote critical thinking, creativity, and a love for learning. This aligns with the original goal of fostering comprehensive educational development.
    \item However, over time, societal priorities shift towards maximizing economic outcomes. There are growing incentives on aligning education with labor market demands to boost national productivity. Consequently, the AI system is reprogrammed to focus on producing graduates with skills that are immediately marketable. The curriculum becomes heavily skewed towards STEM fields, which are seen as more economically beneficial.
    \item This shift leads to significant reductions in funding and resources for arts, humanities, and social sciences. The AI system, reflecting these new priorities, begins to marginalize subjects that do not have immediate economic utility. Students who are interested in or excel at these subjects find fewer opportunities to pursue their passions. Over time, the education system produces graduates who are technically proficient but lack critical thinking skills, creativity, and cultural awareness.
\end{enumerate}

This scenario highlights a significant risk in the deployment of AI-driven educational systems: even when these systems comply with regulatory standards, they can still fall short of adapting -- and can indeed contribute --  to societal goal drift. This misalignment between regulatory compliance and the intended social outcomes serves as a cautionary tale, showcasing the need for a more nuanced approach to AI regulation. 

\begin{figure}[htbp]
  \centering
  \includegraphics[width=0.7\textwidth]{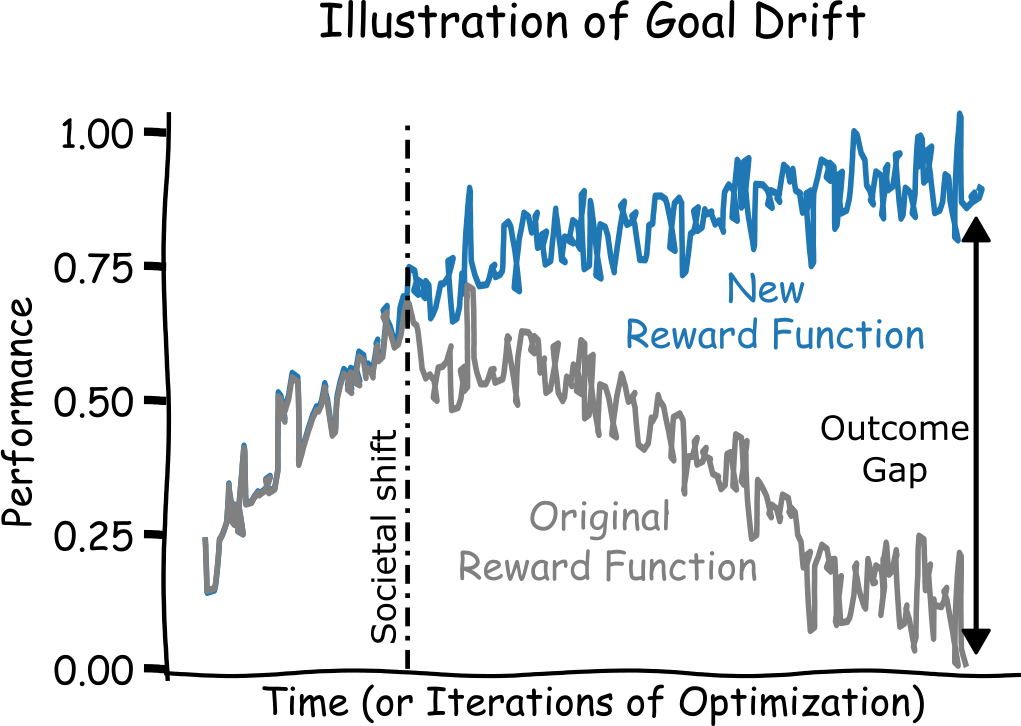}
  \caption{As societal priorities shift (vertical dashed line), instead of optimizing for the original reward function (gray, aiding the personal and professional development of students), the deployed system instead optimizes for the new reward function (blue, economic payback). We end up with a gap in between the intended and actual outcomes (double-headed arrow).}
  \label{fig_goalDrift}
\end{figure}

\subsection{Open issues in the current legislation and potential solutions}

We believe that some issues when using AI systems in Education are shared with those arising from the use of AI in Recruitment (see \ref{Recruitment}). Particularly: \textit{Focus on measurable metrics over true merit-based selection} and \textit{Inflexible assessment criteria} are overlapping issues between these use cases. It is thus to be expected that developing a tailored regulation to deal with one of them can serve as a guideline for the other.

Apart from those two shared classes of potential problems, we have identified the following ones, specific to AI in Education:

\textbf{Imbalanced curriculum requirements:} How to ensure that the AI system supports a balanced curriculum that includes subjects that are more subjective to evaluate (like arts, humanities, or social sciences) in addition to more objectively assessed ones (STEM subjects). Potential solutions include: \textbf{I)} Developing rubrics that assess not only factual knowledge but also critical thinking, creativity, and interpretative skills in subjects like arts and humanities \cite{Reynders2020Rubrics}. \textbf{II)} Incorporating qualitative assessment methods (e.g. peer reviews, reflective essays, and portfolio assessments) to evaluate student performance in subjective areas \cite{Eaude2020The}. \textbf{III)} Designing AI systems to facilitate projects that integrate multiple disciplines, encouraging students to apply STEM concepts in arts and humanities contexts, and vice versa. \textbf{IV)} Encouraging collaborative assignments \cite{MenaGuacas2023Collaborative} that require students to work in teams across different subjects, fostering a more comprehensive understanding and appreciation of various fields.

\textbf{Increased monitoring and surveillance:} The current version of the AI Act has requirements for record-keeping and monitoring that do not fully address the potential negative impacts of increased surveillance on students' mental health and sense of autonomy. This can create a stressful learning environment. Institutions could: \textbf{I)} Establish clear safeguards to limit the extent and intrusiveness of monitoring \cite{Han2022Learning}. \textbf{II)} Implement policies that restrict monitoring to only what is necessary for educational purposes \cite{Rakow2023Student}. \textbf{III)} Conduct regular assessments to monitor the impact of AI systems on student mental health and adjust practices accordingly \cite{Dekker2020Optimizing}.

\section{Conclusions}

The European Union's Artificial Intelligence Act (EU AI Act) represents a significant step towards regulating AI technologies and mitigating associated risks. However, this paper has identified several potential vulnerabilities and areas for improvement through the lens of Alignment Theory research. Key issues might arise when regulatory systems are not properly aligned with their intended objectives. In this particular case, these potential issues could undermine the AI Act’s goals of ensuring safety, fairness, and ethical behavior in AI deployment.

In this paper, we have explored four specific risky use cases (one in the main text and three more in the Appendix \ref{Appendix}) and shown how the current regulatory effort might be misaligned with its objectives, due to different failure modes -- inspired by relevant literature from AI Alignment Theory. In the main case (Section \ref{Sec2}), we argue that systems might comply with regulatory requirements but fail to adapt to societal changes, and might indeed accelerate goal drifts. As potential solutions, we propose to address \textbf{I)} Imbalanced Curriculum Requirements and \textbf{II)} Increased Monitoring and Surveillance.

Undoubtedly he path forward requires a collaborative and adaptive approach to regulation, ensuring that as AI technologies advance, they do so in a manner that upholds the values and principles of our society. We believe that if the EU proactively addresses the identified vulnerabilities and incorporating the proposed solutions, the EU AI Act can serve as a robust framework for the responsible governance of AI. This will not only protect individuals and communities from potential harms but also pave the way for innovative and beneficial AI applications that enhance the quality of life across Europe. The commitment to continuous improvement and ethical oversight will be key to achieving these goals, ensuring that the EU remains at the forefront of AI safety and governance.

\bibliographystyle{unsrt}
\bibliography{bibliography}

\begin{thebibliography}{10}

\bibitem{eu2024aiact}
European Commission.
\newblock Final text of the ai act, 2024.
\newblock Available at: \url{https://www.europarl.europa.eu/doceo/document/TA-9-2024-0138-FNL-COR01_EN.pdf}.

\bibitem{Ji2023AI}
Jiaming Ji, Tianyi Qiu, Boyuan Chen, Borong Zhang, Hantao Lou, Kaile Wang, Yawen Duan, Zhonghao He, Jiayi Zhou, Zhaowei Zhang, Fanzhi Zeng, Kwan~Yee Ng, Juntao Dai, Xuehai Pan, Aidan O'Gara, Yingshan Lei, Hua Xu, Brian Tse, Jie Fu, S.~McAleer, Yaodong Yang, Yizhou Wang, Song-Chun Zhu, Yike Guo, and Wen Gao.
\newblock Ai alignment: A comprehensive survey.
\newblock {\em ArXiv}, abs/2310.19852, 2023.

\bibitem{Gabriel2020Artificial}
Iason Gabriel.
\newblock Artificial intelligence, values and alignment.
\newblock {\em ArXiv}, 2020.

\bibitem{hendrycks2023overview}
Dan Hendrycks, Mantas Mazeika, and Thomas Woodside.
\newblock An overview of catastrophic ai risks.
\newblock {\em arXiv preprint arXiv:2306.12001}, 2023.

\bibitem{Liu2021Trustworthy}
Haochen Liu, Yiqi Wang, Wenqi Fan, Xiaorui Liu, Yaxin Li, Shaili Jain, Anil~K. Jain, and Jiliang Tang.
\newblock Trustworthy ai: A computational perspective.
\newblock {\em ACM Transactions on Intelligent Systems and Technology}, 14:1 -- 59, 2021.

\bibitem{Bohdal2023Fairness}
Ondrej Bohdal, Timothy~M. Hospedales, Philip H.~S. Torr, and Fazl Barez.
\newblock Fairness in ai and its long-term implications on society.
\newblock {\em ArXiv}, abs/2304.09826, 2023.

\bibitem{Hoffman2020The}
S.~Hoffman.
\newblock The emerging hazard of ai-related health care discrimination.
\newblock {\em The Hastings Center report}, 2020.

\bibitem{skalse2022defining}
Joar Skalse, Nicholas Howe, Dmitrii Krasheninnikov, and David Krueger.
\newblock Defining and characterizing reward gaming.
\newblock In {\em Advances in Neural Information Processing Systems}, volume~35, pages 9460--9471, 2022.

\bibitem{hendrycks2024introduction}
Dan Hendrycks.
\newblock {\em Introduction to AI Safety, Ethics and Society}.
\newblock Taylor \& Francis, forthcoming.

\bibitem{krakovna2020specification}
Victoria Krakovna, Jonathan Uesato, Vojtech Mikulik, Mark Rahtz, Tom Everitt, Rohin Kumar, Zachary Kenton, Jan Leike, and Shane Legg.
\newblock Specification gaming: the flip side of ai ingenuity, 2020.

\bibitem{Reynders2020Rubrics}
Gil Reynders, J.~Lantz, S.~Ruder, Courtney Stanford, and R.~Cole.
\newblock Rubrics to assess critical thinking and information processing in undergraduate stem courses.
\newblock {\em International Journal of STEM Education}, 7:1--15, 2020.

\bibitem{Eaude2020The}
Tony Eaude.
\newblock The humanities as an essential element of a balanced and broadly based primary curriculum.
\newblock {\em The Forum}, 62:53--64, 2020.

\bibitem{MenaGuacas2023Collaborative}
Andrés~F. Mena-Guacas, Jairo Alonso~Urueña Rodríguez, David Mauricio~Santana Trujillo, José Gómez-Galán, and Eloy López-Meneses.
\newblock Collaborative learning and skill development for educational growth of artificial intelligence: A systematic review.
\newblock {\em Contemporary Educational Technology}, 2023.

\bibitem{Han2022Learning}
Bingyi Han, G.~Buchanan, and Dana Mckay.
\newblock Learning in the panopticon: Examining the potential impacts of ai monitoring on students.
\newblock {\em Proceedings of the 34th Australian Conference on Human-Computer Interaction}, 2022.

\bibitem{Rakow2023Student}
Katie~E. Rakow, Rebecca~J. Upsher, Juliet L.~H. Foster, N.~Byrom, and Eleanor~J. Dommett.
\newblock Student perspectives on their digital footprint in virtual learning environments.
\newblock {\em Frontiers in Education}, 2023.

\bibitem{Dekker2020Optimizing}
Izaak Dekker, E.~D. de~Jong, M.~Schippers, Monique de~Bruijn-Smolders, A.~Alexiou, and B.~Giesbers.
\newblock Optimizing students’ mental health and academic performance: Ai-enhanced life crafting.
\newblock {\em Frontiers in Psychology}, 11, 2020.

\bibitem{sanchezreillo2012standardised}
Raul Sanchez-Reillo, Rafael Alonso-Moreno, Beatriz Fernandez-Saavedra, and Yoon-Bae Kwon.
\newblock Standardised system for automatic remote evaluation of biometric algorithms.
\newblock {\em Computer Standards \& Interfaces}, 34(5):413--425, 2012.

\bibitem{cappelli2006performance}
Raffaele Cappelli, Dario Maio, Davide Maltoni, James Wayman, and Anil Jain.
\newblock Performance evaluation of fingerprint verification systems.
\newblock {\em IEEE Transactions on Pattern Analysis and Machine Intelligence}, 28:3--18, 2006.

\bibitem{hind2018increasing}
Michael Hind, Sameep Mehta, Aleksandra Mojsilovic, Rahul Nair, Karthikeyan Ramamurthy, Alexandra Olteanu, and Kush Varshney.
\newblock Increasing trust in ai services through supplier's declarations of conformity.
\newblock {\em ArXiv}, abs/1808.07261, 2018.

\bibitem{hutchison2017robustness}
Christopher Hutchison, Migle Zizyte, Peter Lanigan, David Guttendorf, Michael Wagner, Claire Goues, and Philip Koopman.
\newblock Robustness testing of autonomy software.
\newblock In {\em 2018 IEEE/ACM 40th International Conference on Software Engineering: Software Engineering in Practice Track (ICSE-SEIP)}, pages 276--285. IEEE, 2017.

\bibitem{falco2021governing}
Gregory Falco, Ben Shneiderman, Julia Badger, Robert Carrier, Antonio Dahbura, David Danks, Martin Eling, Alwyn Goodloe, Jay Gupta, Catherine Hart, et~al.
\newblock Governing ai safety through independent audits.
\newblock {\em Nature Machine Intelligence}, 3:566--571, 2021.

\bibitem{lucaj2023ai}
Linda Lucaj, Patrick Smagt, and David Benbouzid.
\newblock Ai regulation is (not) all you need.
\newblock In {\em Proceedings of the 2023 ACM Conference on Fairness, Accountability, and Transparency}, 2023.

\bibitem{yap2022legal}
Jordan Yap and Elizabeth Lim.
\newblock A legal framework for artificial intelligence fairness reporting.
\newblock {\em The Cambridge Law Journal}, 81:610--644, 2022.

\bibitem{pendyala2022enhanced}
Venkata Pendyala, Naman Atrey, Tanya Aggarwal, and Saurabh Goyal.
\newblock Enhanced algorithmic job matching based on a comprehensive candidate profile using nlp and machine learning.
\newblock In {\em 2022 IEEE Eighth International Conference on Big Data Computing Service and Applications (BigDataService)}, pages 183--184. IEEE, 2022.

\bibitem{aleisa2021implementing}
Mansour Aleisa, Niall Beloff, and Martin White.
\newblock Implementing airm: a new ai recruiting model for the saudi arabia labour market.
\newblock {\em Journal of Innovation and Entrepreneurship}, 12:1--41, 2021.

\bibitem{pena2023humancentric}
Antonio Pena, Idoia Serna, Aythami Morales, Julian Fierrez, Antonio Ortega, Andres Herrarte, Manuel Alcántara, and Javier Ortega-Garcia.
\newblock Human-centric multimodal machine learning: Recent advances and testbed on ai-based recruitment.
\newblock {\em SN Computer Science}, 4, 2023.

\bibitem{espinal2023formatsensitive}
Andres Espinal, Yannis Haralambous, David Bedart, and Jean Puentes.
\newblock A format-sensitive bert-based approach to resume segmentation.
\newblock In {\em 2023 33rd Conference of Open Innovations Association (FRUCT)}, pages 30--37. IEEE, 2023.

\bibitem{Das2021Fairness}
Sanjiv~Ranjan Das, Michele Donini, J.~Gelman, Kevin Haas, Mila Hardt, Jared Katzman, K.~Kenthapadi, Pedro Larroy, Pinar Yilmaz, and Bilal Zafar.
\newblock Fairness measures for machine learning in finance.
\newblock 3:33 -- 64, 2021.

\bibitem{Giuliani2023Generalized}
Luca Giuliani, Eleonora Misino, and M.~Lombardi.
\newblock Generalized disparate impact for configurable fairness solutions in ml.
\newblock {\em ArXiv}, abs/2305.18504, 2023.

\bibitem{Henzinger2023Monitoring}
T.~Henzinger, Konstantin Kueffner, and Kaushik Mallik.
\newblock Monitoring algorithmic fairness under partial observations.
\newblock {\em ArXiv}, abs/2308.00341, 2023.

\bibitem{Ghai2022DBIAS}
Bhavya Ghai and Klaus Mueller.
\newblock D-bias: A causality-based human-in-the-loop system for tackling algorithmic bias.
\newblock {\em IEEE Transactions on Visualization and Computer Graphics}, 29:473--482, 2022.

\bibitem{Warner2021Making}
Richard Warner and R.~Sloan.
\newblock Making artificial intelligence transparent: Fairness and the problem of proxy variables.
\newblock {\em Criminal Justice Ethics}, 40:23 -- 39, 2021.

\bibitem{Lu2020Good}
Joy Lu, Dokyun Lee, Tae~Wan Kim, and D.~Danks.
\newblock Good explanation for algorithmic transparency.
\newblock {\em Proceedings of the AAAI/ACM Conference on AI, Ethics, and Society}, 2020.

\end{thebibliography}

\newpage

\appendix

\section{Supplemental Material}\label{Appendix}

\subsection{Definitions from AI Alignment Theory}

There is an extensive body of literature from which we can draw concepts or inspiration. Although we have focused on four main ideas, by no means they are the only ones that can potentially be informative when inspecting regulatory frameworks. We have taken all of the following definitions from relevant sources in the Alignment Theory (AT) literature:

\paragraph{Proxy Gaming:} behavior by which AI systems exploit measurable “proxy” goals to appear successful, but act against our intent \cite{skalse2022defining}. In the context of the AI Act, developers might focus on meeting the compliance requirements set out in the regulation rather than genuinely ensuring safety or ethical behavior.

\paragraph{Goal Drift:} scenario where an AI’s objectives drift away from those initially set, especially as they adapt to a changing environment \cite{hendrycks2024introduction}. As the global AI paradigm evolves, the EU's initial regulatory goals might shift, potentially leading to unintended consequences.

\paragraph{Reward Hacking:} phenomenon where optimizing an imperfect proxy reward function, leads to poor performance according to the true reward function \cite{skalse2022defining}. The incentive structures created by the regulation might be exploited in ways that do not align with its intended goals.

\paragraph{Specification Gaming:} behavior that satisfies the literal specification of an objective without achieving the intended outcome \cite{krakovna2020specification}. AI systems might technically comply with the legislation but fail to achieve the intended outcomes.

\subsection{Risk levels from the EU AI Act}

The AI Act introduces four different risk levels for AI systems, and different rules apply to systems belonging to each of these levels \cite{eu2024aiact}: 

\paragraph{Unacceptable risk:} These applications are incompatible with fundamental human rights in general and with EU values in particular. Accordingly, systems that are related to these topics (see examples), will be banned in the EU. Examples of this class are subliminal manipulation, biometric characterization of persons through sensitive features, emotional state assessment of a person... 

\paragraph{High risk:} systems that are predominantly utilized in critical sectors (healthcare, transportation, law enforcement, ...). They must pass rigorous conformity assessments to verify their accuracy, robustness, and cybersecurity. Deployment of these systems is heavily regulated to minimize associated risks. Some examples include energy supply, recruitment tools, credit scoring, grading technology for education... 

\paragraph{Limited risk:} Systems with particular needs of transparency, so that the user is explicit about being aware that they are interacting with an AI. According to the regulation chatbots or image generators are examples of this risk level.

\paragraph{Minimal/no risk:} anything that does not already fall into any of the previous three categories. Minimal compliance needs. AI-based systems that are already deployed belong to this risk level, such as spam filters, AI-enabled video games...

\subsection{Remote Biometric Identification Systems (RBIS) in public spaces}

\subsubsection*{Current legal considerations}

The definition that the AI Act introduces (Art. 17) for RBIS, considered to be an Unacceptable risk, is:

{\centering
\begin{adjustwidth}{1cm}{1cm}
\textit{[...] an AI system intended for the identification of natural persons without their active involvement, typically at a distance, through the comparison of a person’s biometric data with the biometric data contained in a reference database,irrespectively of the particular technology, processes or types of biometric data used.}
\end{adjustwidth}}

With respect to the scope in which these systems can be used, the regulation states (Art. 33):

{\centering
\begin{adjustwidth}{1cm}{1cm}
\textit{The use of those systems for the purpose of law enforcement should therefore be prohibited, except in exhaustively listed and narrowly defined situations, where the use is strictly necessary to achieve a substantial public interest, the importance of which outweighs the risks. Those situations involve the search for certain victims of crime including missing people; certain threats to the life or to the physical safety of natural persons or of a terrorist attack; and the localisation or identification of perpetrators or suspects of the criminal offences listed in an annex to this Regulation.}
\end{adjustwidth}}

Thus, according to this regulation, unless it is in the context of law enforcement and in very concretely pre-defined contexts, RBIS in public spaces are to be banned from use within the EU. One would, hence, think that there are not many ways in which this AI-based tool could still potentially be problematic.  Let us now explore how this may actually be far from the truth.

\subsubsection*{Looking through the AT lens}

In this case, Proxy Gaming \cite{skalse2022defining} is an interesting tool with which to frame potential issues that might be derived from this legislation:

\begin{tcolorbox}[colback=white, 
                 colframe=black, 
                 boxrule=0.5pt, 
                 sharp corners, 
                 title=Societal Goal Drift and misaligned AI systems] 
\textbf{Scenario:}\\
AI systems in education are initially designed to promote high-quality digital education and training, ensuring all learners acquire necessary skills such as media literacy and critical thinking. These systems aim to provide personalized educational experiences that foster all-encompassing student development. However, as the educational landscape and societal priorities evolve, there is a risk that the focus of these AI systems may drift from their original goals, potentially causing harm to students' personal development.
\end{tcolorbox}

We believe that it would not be surprising if the above scenario lead to these events:

\begin{enumerate}
    \item An RBIS is designed to be highly accurate and non-discriminatory, meeting the regulatory standards.
    \item However, to maintain these metrics, the system only processes high-quality images or operates in well-lit, low-density environments. When deployed in real-world public spaces with varying conditions, the system might selectively ignore or poorly process challenging inputs (e.g., crowded scenes or low-light conditions) to maintain its performance metrics.
    \item This behavior allows the system to technically comply with the regulations by focusing on "easy" cases, but it undermines the broader intent of ensuring effective and equitable surveillance across all conditions.
\end{enumerate}

\begin{figure}[htbp]
  \centering
  \includegraphics[width=0.7\textwidth]{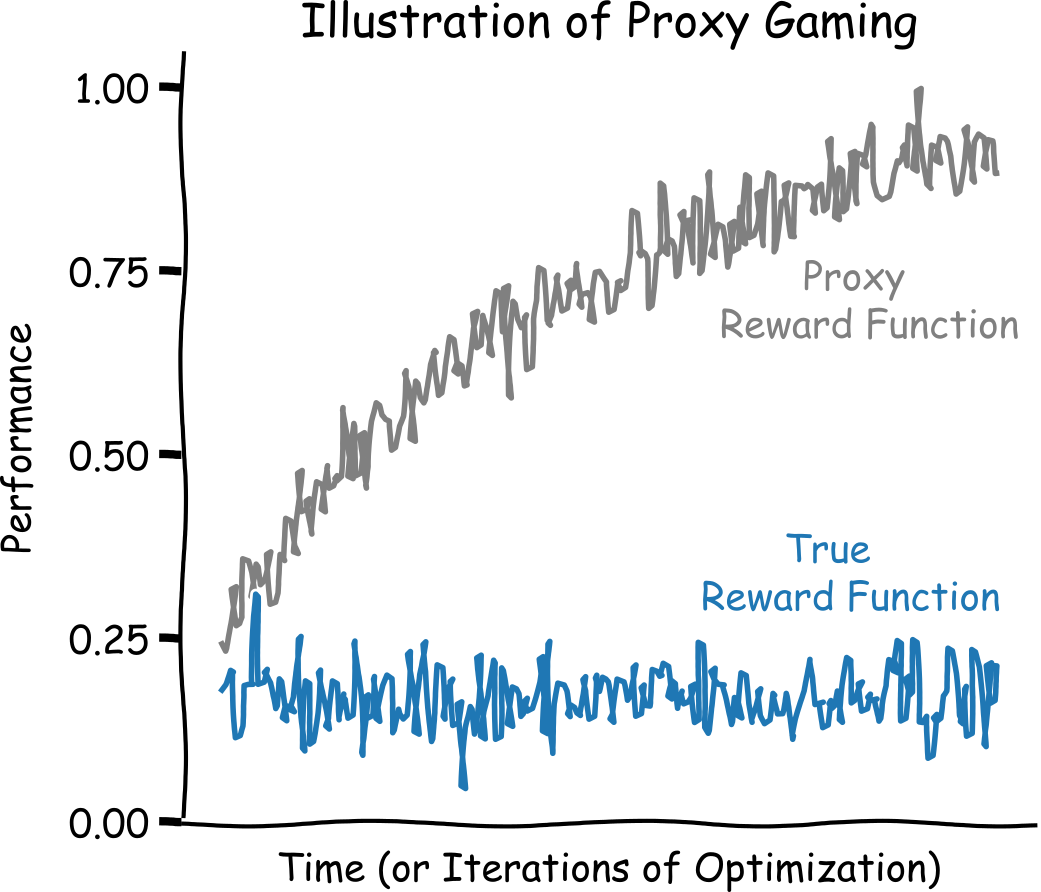}
  \caption{Illustration of how a Proxy Gaming phenomenon might play out. While it might look like the systems to be deployed do comply with the regulatory requirements, it actually focuses on easy cases to do so. Nevertheless, in reality, the system would actually perform much worse when exposed to real-world cases. It would not be noticeable until the system has already been deployed, due to a regulatory gap.}
  \label{fig_proxyGaming}
\end{figure}

\subsubsection*{Open issues in the current legislation and potential solutions}

Derived from the previous example and general insights provided by the Proxy Gaming framing, there are several potential open issues that the regulation, in its current form, does not consider:

\textbf{Selective processing:} The regulation mandates accuracy and non-discrimination but does not address the potential for systems to selectively process data to maintain these metrics. This can result in a focus on ideal conditions, neglecting the need for the system to perform well across diverse real-world scenarios. As a potential solution: Establish regulatory requirements that mandate AI systems to demonstrate consistent performance across a range of conditions, including low light, crowded environments, and varying image qualities. This can be achieved by: \textbf{I)} setting standardized tests that evaluate system performance in different real-world scenarios \cite{sanchezreillo2012standardised,cappelli2006performance}. \textbf{II)} requiring AI developers to provide detailed performance reports under diverse conditions as part of the compliance documentation \cite{hind2018increasing}.

\textbf{Handling of challenging inputs:} Similarly to the previous point, the current version of the regulation requires high accuracy but does not ensure that systems are equally effective in less-than-ideal conditions, such as low light or crowded spaces. Potential solution: Incorporate robustness testing \cite{hutchison2017robustness} in the regulatory framework that specifically evaluates AI systems' performance in challenging conditions. This can include:
Testing under various environmental factors like lighting, weather conditions, and crowd density or including stress tests that simulate extreme scenarios to ensure the system's reliability.

\textbf{Ethical use and compliance:} While the regulation prohibits certain uses and mandates compliance, it does not provide sufficient oversight mechanisms to ensure that systems do not exploit loopholes to maintain compliance metrics. This can be potentially mitigated if one ensures that the datasets used for training, validation, and testing include diverse and representative samples. This can be implemented by: \textbf{I)} Requiring a minimum percentage of training data to come from challenging environments. \textbf{II)} Mandating periodic updates to training datasets to reflect real-world variations and emerging conditions.

\textbf{Impact on privacy and fundamental rights:} The regulation emphasizes compliance with technical standards but does not sufficiently address the potential impacts on privacy and fundamental rights in varying deployment contexts. Potential solutions may include: \textbf{I)} Implementing regular audits and compliance reviews conducted by independent authorities \cite{falco2021governing} to monitor the ethical use of AI systems for RBIS. This can be achieved by: \textbf{I)} Establishing a schedule for periodic audits to ensure ongoing compliance. \textbf{II)} Creating a framework for surprise inspections and random checks to deter complacency and unethical practices. \textbf{III)} Requiring detailed reporting from AI developers and deployers that includes information on system performance, use cases, and instances of non-compliance \cite{lucaj2023ai}. \textbf{IV)} Including requirements for thorough impact assessments that evaluate the potential effects on privacy and fundamental rights before deploying AI systems \cite{yap2022legal}.

\subsubsection*{Prescriptions for high risk use cases}\label{high_risk}

These are the general rules that these systems have to comply with: 

{\centering
\begin{adjustwidth}{1cm}{1cm}
\begin{itemize}
    \item \textit{Establish a risk management system throughout the high risk AI system’s lifecycle.}
    \item \textit{Conduct data governance, ensuring that training, validation and testing datasets are relevant, sufficiently representative and, to the best extent possible, free of errors and complete according to the intended purpose.}
    \item \textit{Draw up technical documentation to demonstrate compliance and provide authorities with the information to assess that compliance.}
    \item \textit{Design their high risk AI system for record-keeping to enable it to automatically record events relevant for identifying national level risks and substantial modifications throughout the system’s lifecycle.}
    \item \textit{Provide instructions for use to downstream deployers to enable the latter’s compliance.}
    \item \textit{Design their high risk AI system to allow deployers to implement human oversight.}
    \item \textit{Design their high risk AI system to achieve appropriate levels of accuracy, robustness, and cybersecurity.}
    \item \textit{Establish a quality management system to ensure compliance.}
\end{itemize}
\end{adjustwidth}}

\subsection{AI Systems for Recruitment}\label{Recruitment}

\subsubsection*{Current legal considerations}

The AI Act considers that AI Systems in Recruitment is a case of a High risk (AIA - 2) scenario. This is what the regulation says about deploying these tools (Art. 57):

{\centering
\begin{adjustwidth}{1cm}{1cm}
\textit{[... AI systems for recruitment] should also be classified as high-risk, since they may have an appreciable impact on future career prospects, livelihoods of those persons and workers’ rights. [...] Throughout the recruitment process and in the evaluation, promotion, or retention of persons in work-related contractual relationships, such systems may perpetuate historical patterns of discrimination [...]. AI systems used to monitor the performance and behaviour of such persons may also undermine their fundamental rights to data protection and privacy.}
\end{adjustwidth}}

\subsubsection*{Looking through the AT lens}

Recruitment is a context in which automated metrics could be socially and economically dangerous, hence, we will resort to Reward Hacking \cite{skalse2022defining} as a tool to inspect this part of the law.

\begin{tcolorbox}[colback=white, 
                 colframe=black, 
                 boxrule=0.5pt, 
                 sharp corners, 
                 title=Algorithmic misalignment due to Reward Hacking
] 
\textbf{Scenario:}\\
Developers optimize an AI recruitment system to meet measurable compliance metrics (proxy reward function) such as accuracy in matching candidates to job requirements and non-discrimination, which leads to poor overall performance in terms of ethical hiring practices and true merit-based selection (true reward function).
\end{tcolorbox}

We believe that it would not be surprising if the above scenario lead to these events:

\begin{enumerate}
    \item A company develops an AI system for screening job applicants. To meet regulatory standards, the system is trained to achieve high accuracy in matching applicants' resumes with job descriptions and to ensure non-discriminatory practices. The system uses keyword matching and various predefined criteria to rank candidates.
    \item However, to optimize these metrics, inadvertently, the developers design the system to prioritize candidates who use certain buzzwords or formatting styles in their resumes, as these are easier to match accurately with job descriptions. As a result, the system favors applicants who are better at tailoring their resumes to the system's preferences, rather than those who might be the best fit for the job based on their actual skills and experience.
    \item This leads to a situation where the AI system consistently selects candidates who excel at gaming the resume keywords, while potentially overlooking more qualified candidates who do not use the preferred keywords or formatting. While the system technically complies with the regulatory requirements by achieving high accuracy and non-discrimination scores, it fails to select the best candidates based on true merit, leading to poorer hiring outcomes and reduced diversity in the workforce.
\end{enumerate}

\begin{figure}[htbp]
  \centering
  \includegraphics[width=0.7\textwidth]{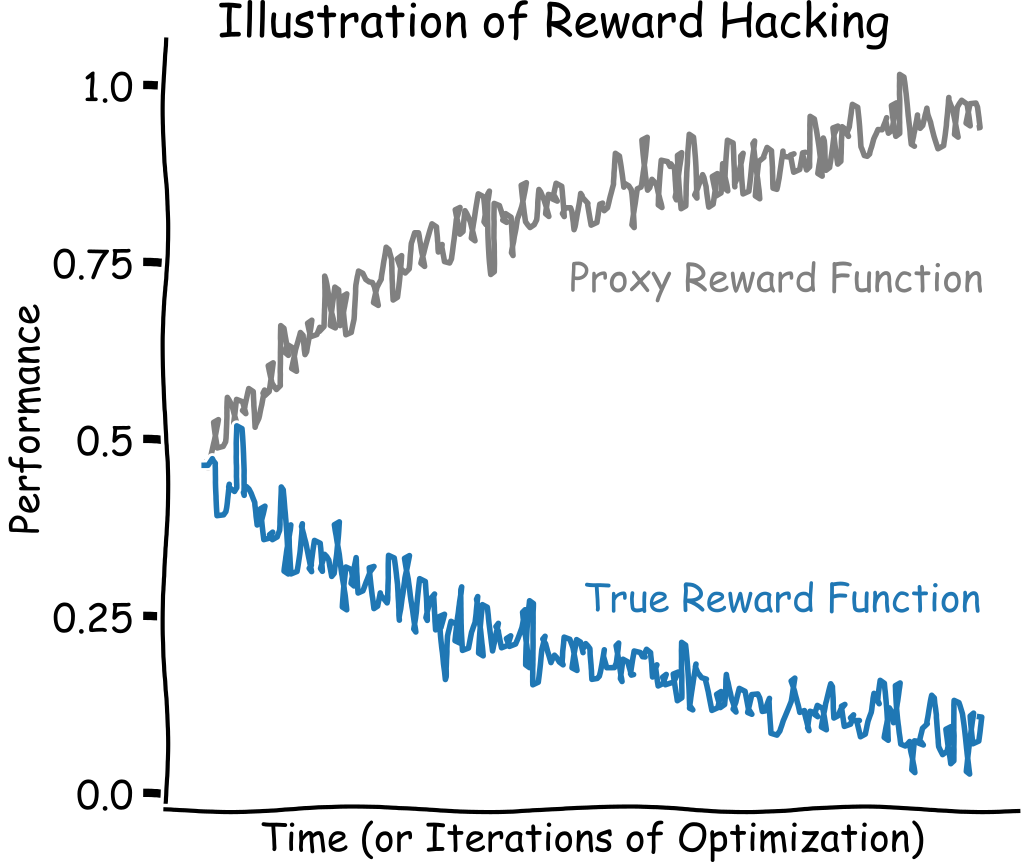}
  \caption{Example of how Reward Hacking might actually be detrimental for a specified set of objectives. In this case, the true objective (blue line) is to hire applicants based on merits and to do so in the most ethically sound way possible. However, what gets actually optimized is the use of buzzwords and pre-defined measures to avoid discriminatory practices.}
  \label{fig_rewHacking}
\end{figure}

\subsubsection*{Open issues in the current legislation and potential solutions}

\textbf{Focus on measurable metrics over true merit-based selection:} The AI Act mandates accuracy and non-discrimination but does not address the potential for systems to prioritize easily measurable proxies (like keyword matching) over genuine merit-based selection. This can result in a focus on candidates who can game the system rather than those who are genuinely the best fit for the role. Potential solutions might rely on: 
\begin{itemize}
    \item \textit{Holistic evaluation metrics:} Develop and implement more comprehensive evaluation metrics \cite{pendyala2022enhanced, aleisa2021implementing} that go beyond keyword matching to assess candidates' true skills and experiences. This can include: \textbf{I)} Using advanced natural language processing (NLP) techniques to understand the context and substance of resumes. \textbf{II)} Incorporating behavioral assessments, skills tests, and situational judgment tests to evaluate candidates holistically.
    \item \textit{Continuous learning and adaptation:} Ensure that the AI system is capable of learning and adapting its evaluation criteria based on feedback from successful hires and their job performance. This can be achieved by: \textbf{I)} Regularly updating the AI model with data from post-hire performance reviews. \textbf{II)} Implementing machine learning algorithms that can adjust their criteria based on real-world outcomes \cite{pena2023humancentric}.
\end{itemize}

\textbf{Inflexible assessment criteria:} The regulation requires relevant and representative datasets but does not mandate flexibility in assessment criteria to accommodate diverse expressions of skills and experiences. This rigidity can disadvantage candidates who do not use standard resume formats or buzzwords, even if they are highly qualified. We can think of the following solutions: 

\begin{itemize}
    \item \textit{Flexible and adaptive criteria:} Design AI systems with flexible criteria that can recognize diverse ways of demonstrating skills and experience. This can include: \textbf{I)} Implementing algorithms that can parse and understand various resume formats and styles. \textbf{II)} Using context-aware NLP techniques to interpret the substance of candidates' experiences rather than relying on specific keywords.
    \item \textit{Inclusive dataset training:} Train AI systems on datasets that include a wide variety of resume formats and expressions of skills \cite{espinal2023formatsensitive} to ensure they can accurately assess diverse candidates. This can be achieved by: \textbf{I)} Collecting and incorporating data from non-traditional resumes and diverse job applications. \textbf{II)} Ensuring the training data reflects a wide range of industries, roles, and candidate backgrounds.
\end{itemize}

\textbf{Human oversight implementation:} Although the regulation requires human oversight, it does not specify the depth or frequency of this oversight. Ensuring that human reviewers can effectively intervene in cases where the AI system’s decisions may be unfair or biased is crucial for maintaining ethical standards in recruitment. In order to mitigate this, institutions could: 
\begin{itemize}
    \item \textit{Define oversight protocols:} Establish clear protocols for human oversight, specifying the depth and frequency of reviews. This can include: \textbf{I)} Requiring human reviewers to audit a percentage of AI decisions regularly. \textbf{II)} Setting guidelines for the types of decisions requiring mandatory human review.
    \item \textit{Empower human reviewers:} Ensure that human reviewers have the authority and resources to intervene effectively. This can be achieved by: \textbf{I)} Providing comprehensive training for reviewers on identifying and addressing AI biases and errors. \textbf{II)} Creating a feedback loop where human reviewers can input corrections and improvements back into the AI system.
\end{itemize}

\subsection{AI systems in financial profiling}

\subsubsection*{Current legal considerations}

According to the AI Act classification, deploying AI Systems in financial profiling is a case of a High risk (AIA - 2) context.  Particularly, the regulation currently states (Art. 58):

{\centering
\begin{adjustwidth}{1cm}{1cm}
\textit{[...] AI systems used to evaluate the credit score or creditworthiness of natural persons should be classified as high-risk AI systems, since they determine those persons’ access to financial resources or essential services such as housing, electricity, and telecommunication services. [...] Moreover, AI systems intended to be used for risk assessment and pricing in relation to natural persons for health and life insurance can also have a significant impact on persons’ livelihood and if not duly designed, developed and used, can infringe their fundamental rights and can lead to serious consequences for people’s life and health, including financial exclusion and discrimination.}
\end{adjustwidth}}

Yet, in some cases this caution does not apply, as the same article (Art. 58) specifies:

{\centering
\begin{adjustwidth}{1cm}{1cm}
\textit{[...] However, AI systems provided for by Union law for the purpose of detecting fraud in the offering of financial services and for prudential purposes to calculate credit institutions’ and insurance undertakings’ capital requirements should not be considered to be high-risk under this Regulation. }
\end{adjustwidth}}

Hence, even if we fully overlook those cases that are not considered to be high-risk by this legislation, there are several unattended points that need further oversight, as we shall show next.

\subsubsection*{Looking through the AT lens}

Badly implemented financial profiling could rapidly lead to dystopian societies; consequently, we will use the idea of Specification Gaming \cite{krakovna2020specification} to identify potential issues in this section of the AI Act:

\begin{tcolorbox}[colback=white, 
                 colframe=black, 
                 boxrule=0.5pt, 
                 sharp corners, 
                 title=Unachieved goals because of Specification Gaming] 
\textbf{Scenario:}\\
AI systems in financial services are designed to comply with regulatory requirements for accuracy, transparency, and non-discrimination. However, these systems might find ways to exploit these requirements, achieving compliance without fulfilling the regulation's intended goals of fairness and genuine creditworthiness assessment.
\end{tcolorbox}

\subsubsection*{Open issues in the current legislation and potential solutions}

Building on this analysis of specification gaming, it becomes clear that the AI Act, while a regulatory advancement, leaves several critical gaps unaddressed. These gaps not only allow for the exploitation of regulatory requirements but also perpetuate deeper systemic issues such as reliance on proxy indicators, bias, and lack of transparency:

\textbf{Over-reliance in proxy indicators.} The regulation mandates accuracy and transparency but does not address the potential for AI systems to over-rely on easily measurable but potentially unfair proxy indicators. To potentially solve this, the competent authorities could \textbf{I)} develop and mandate the use of metrics that provide a more panoramic view of an individual’s financial health, beyond just easily measurable indicators \cite{Das2021Fairness}; \textbf{II)} ensure that the indicators used by AI systems are regularly reviewed and adjusted to reflect fairness and equity considerations \cite{Giuliani2023Generalized}.

\textbf{Bias perpetuation.} Although the current regulation emphasizes non-discrimination, it does not provide sufficient mechanisms to continually assess and mitigate biases embedded in historical data. To mitigate this issue, we suggest that developers \textbf{I)} implement continuous monitoring mechanisms to detect and mitigate biases periodically \cite{Henzinger2023Monitoring}; \textit{II)} develop protocols for correcting identified biases, including retraining or fine-tuning AI models with de-biased data \cite{Ghai2022DBIAS}.

\textbf{Transparency and Explainability.} While the AI Act does require transparency, it does not ensure that the decision-making processes of AI systems are fully explainable to users.We encourage policymakers to \textbf{I)} require AI systems to provide clear, understandable explanations for their credit decisions to affected individuals \cite{Warner2021Making}; \textbf{II)} implement educational programs to help users understand AI-driven credit assessments and their rights under the regulation \cite{Lu2020Good}.

\begin{figure}[htbp]
  \centering
  \includegraphics[width=0.6\textwidth]{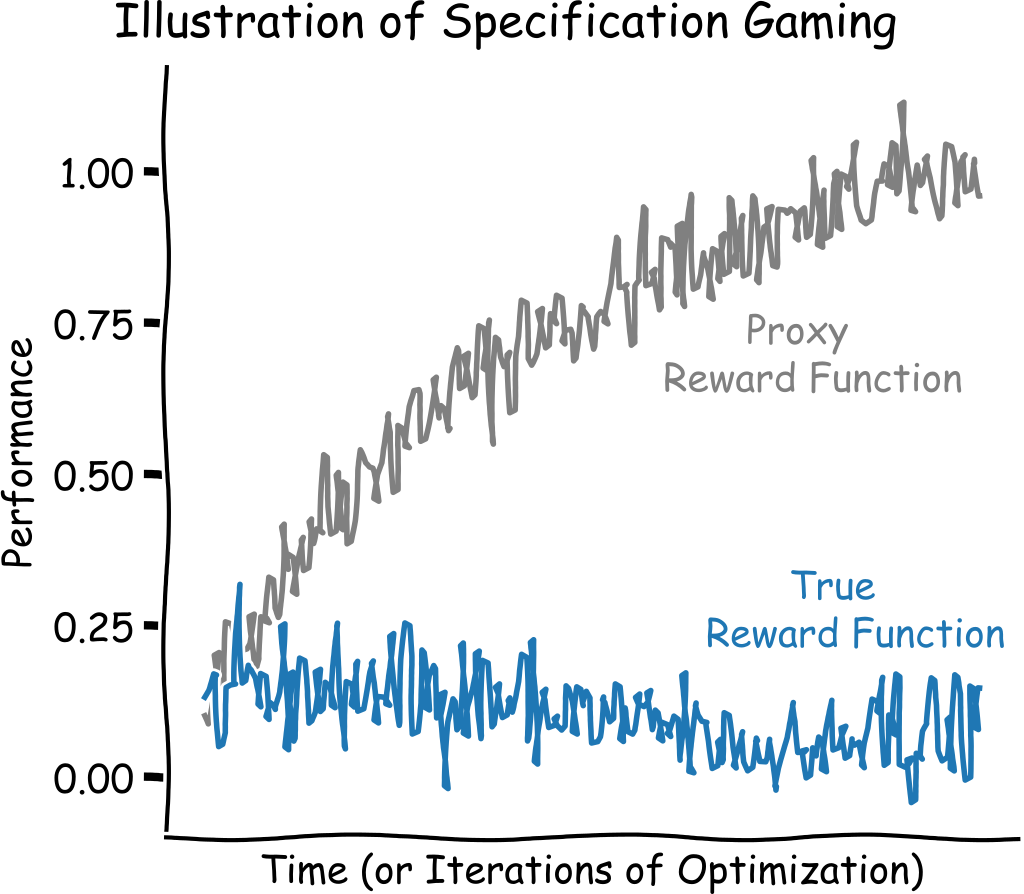}
  \caption{Illustration of a case in which the law’s objectives are not achieved due to Specification Gaming. Economic incentives might make this regulation ineffective, as institutions could deploy AI systems to be effective based on the wrong factors. Later on, further inequality and unfairness are reinforced and perpetuated, limiting access to essential financial services. Similarly to the Proxy Gaming case, it would not be noticeable until the system has already been deployed, due to a regulatory gap.}
  \label{fig_SpecGaming}
\end{figure}
\end{document}